\documentclass[preprint,prb,showpacs]{revtex4}
\begin{document}
\topmargin-1.0cm
\ifx\undefined\psfig\def\psfig#1{    }\else\fi

\title  {All-optical single-electron read-out devices based on GaN quantum dots}
\author  {Irene D'Amico$^{1,2}$, Fausto Rossi$^{1,2,3}$}
\affiliation{ $^{1}$Istituto Nazionale per la Fisica della Materia (INFM);\\
 $^{2}$Institute for Scientific Interchange, via Settimio Severo 65, 
I-10133 Torino,
Italy; \\$^{3}$Dipartimento di Fisica, Politecnico di Torino, Corso Duca degli 
Abruzzi 24, I-10129 Torino,
Italy}
\date{\today}  
\begin{abstract}
We study few-particle interactions in GaN-coupled quantum dots and discuss how
the built-in field characteristic of these structures strongly reinforce
dipole-dipole and dipole-monopole interactions.
 We introduce a semi-analytical model which allows for a rapid and easy 
estimate of the magnitude of few-particle interactions and whose predictions 
are
 closer
than 10\% to ``exact'' results. We apply our study to the design of an
 all-optical read-out device which exploits long-range dipole-monopole 
interactions and may be also used to monitor the charge status of a quantum dot
 system.
\end{abstract}
\maketitle
Quantum dots are quasi-zero dimensional systems which have focused
 the attention of the scientific community as a playground for testing
few-particle fundamental physics  and for
 their increasing technological applications.\cite{dot,theotechno,Zrenner,PRLBiolatti,PRBBiolatti}
 In particular semiconductor self-assembled quantum dots have received large 
attention  recently due to the symmetry of their confinement potential (which paves the road to an easier engeenering of their electronic structure) and to  the possibility of optically-driving their 
excitations; this in turn allows for ultrafast state
 manipulation and coeherent control
 of the few particle 
dynamics, a basic request for quantum computing devices.\cite{PRLBiolatti,PRBBiolatti}
In this panorama
 GaN self-assembled quantum dots (QDs) occupy a distinguished place
 since, while
 their characterization is still incomplete, 
 interesting properties such as their long spin decoherence time\cite{Awscha} 
or the presence of a  built-in electric field ---as 
strong as 
few MeV/cm\cite{GaNgeneral}--- have been already demonstrated.
 Such a field,  of polarization and piezoelectric origin,  is antiparallel to
the growth direction inside the QDs, inverting its sign outside.\cite{GaNgeneral}
Its importance in relation to single-electron devices 
lays  on the possibility of creating 
 strong bonds between neighboring quantum dots which can be switched on/off 
optically.  Under its action in fact,
 electrons and holes are 
driven in opposite directions, so to create electrical dipoles which interacts
with energies of the order of few meV;\cite{DeRinaldis1}
this effect is, for example, the key ingredient for a recently proposed 
quantum computation/information processing device.\cite{DeRinaldis1}

In this Letter we shall (i)provide a very accurate
 semi-analytical method to estimate
 few particle  interactions between stacked 
QDs without resorting to heavy numerical calculations and  
(ii)propose an all-optical read-out device based on such interactions.

Let us consider first a single exciton confined in a GaN QD.
 We shall work in the usual envelope function approximation and focus on the system ground state, though our method 
can be easily extended to the other low-energy level states.
The Hamiltonian of the excitonic system is
$H=\sum_{\alpha=e,h}\left[H^0_{z\alpha}+m_\alpha\omega_\alpha^2(x_\alpha^2+y_\alpha^2)/2
\right]-
{e^2/ \epsilon |\vec{r_e}-\vec{r_h}|}$
with $H_{z (e/h)}^0=(p_{z(e/h)}^2/2m_{e/h})+V_z(z_{e/h})\pm eE(z)z_{e/h}$. 
The terms in square bracket correspond to the Hamiltonian of a particle 
confined by a parabolic potential in the in-plane directions and by the 
strongly confining potentials $V_z$ in the growth direction (usually modeled as a square potential), $E(z)$ is the built-in electric field, 
 $e$ the absolute value of the electron charge and $\epsilon$ the dielectric constant of the medium. In this paper 
 Greek letters will indicate the indexes   $e,h$ corresponding 
respectively to electrons and holes.

The ground state $\psi_{e/h}(z_{e/h})$ of the one-dimensional Hamiltonian
$H_{z (e/h)}^0$, can be easily 
calculated by exact diagonalization.
We can then resort to the  separable 
effective  Hamiltonian $\tilde{H}= H_{ze}^0+H_{zh}^0+H_I$ where
\begin{eqnarray}
& & H_I(\vec{R},\vec{r})= \left[{p_R^2\over2M}+{1\over2}M\omega_R^2R^2\right]
+\left[{p_r^2\over2\mu}+{1\over2}\mu\omega_r^2r^2\right.\nonumber \\&-&\left.{e^2\over \epsilon\sqrt{r^2+\langle (z_e-z_h)^2\rangle}}\right] 
+\mu(\omega_e^2-\omega_h^2)\vec{R}\cdot\vec{r}.\label{H_I}
\end{eqnarray}
Here
$\vec{R}=[m_e(x_e,y_e)+m_h(x_h,y_h)]/ M$ and
$\vec{r}=(x_e-x_h,y_e-y_h)$ are the in-plane center of mass and relative coordinates, $M= m_e+m_h$, $\mu= m_em_h/M$,
$\omega_R^2=(m_e\omega_e^2+m_h\omega_h^2)/ M$,
$\omega_r^2=(m_h\omega_e^2+m_e\omega_h^2)/ M$
and $\langle (z_e-z_h)^2\rangle\equiv \langle\psi_e(z_e)\psi_h(z_h)|(z_e-z_h)^2
|\psi_e(z_e)\psi_h(z_h)\rangle$.
The original problem has now been reduced to solving the
Schroedinger equation for $H_I(x,y)$.
By approximating the ground state solution of $H_I$ with the factorized form
$\psi_R(\vec{R})\psi_{rel}(\vec{r})$, where
$\psi_R(\vec{R})=\sqrt{M\omega_R/\hbar\pi}\exp(-M\omega_RR^2/2\hbar)$,
we get
\begin{eqnarray}
\tilde{H}_{rel}(r)&=&\langle\psi_R|H_I|\psi_R\rangle
=\hbar\omega_R+{p_r^2\over2\mu}+V(r)\label{H_rel}
\end{eqnarray}
with
$V(r)={1\over2}\mu\omega_r^2r^2-{e^2/ \epsilon\sqrt{r^2+\langle 
(z_e-z_h)^2\rangle}}$.
For calculating the properties of 
 low-energy states, $V(r)$ can be approximated around its minimum as
$V(r)\approx V_0+\mu\tilde{\omega}_r^2r^2/2$
where
\begin{equation}
\mu\tilde{\omega}_r^2=\mu\omega_r^2+{e^2\over \epsilon\langle(z_e-z_h)^2\rangle^{3/2}}.\label{om_til}
\end{equation} We underline that the expression (\ref{om_til}) includes
  corrections due to the Coulomb interaction between electron and hole. As we will see later, such corrections strongly improve the precision of the 
approximation.\cite{wfz}
 The eigenvalue problem related to Eq.~(\ref{H_rel})
is now exactly solvable and its 
ground state  is given by 
$\psi_{rel}(r)=\sqrt{\mu\tilde{\omega}_r/\hbar\pi}\exp(-\mu\tilde{\omega}_rr^2/2\hbar)$. 
The approximated form  for the total excitonic wavefunction is then 
\begin{equation}
\psi_{xc}\approx\psi_e(z_e)\psi_h(z_h)\psi_R(R)\psi_{rel}(r).
\label{excwf}
\end{equation}

Next we will discuss the bi-exciton system.
If the barrier between two stacked 
quantum dots QD0 and QD1 is such that particle 
tunneling is negligible, or if the mismatch between 
relevant single particle levels in  QD0 and QD1 is sufficiently large due to
 the built-in electric field and to size differences between the two dots, each direct low energy  exciton
in the macromolecule QD0 + QD1 will be strictly confined to a single dot,
 so that when considering a biexciton formed by one exciton in QD0
and the second in QD1,
we can safely approximate its wavefunction as the product 
$\psi_{bi}\approx\psi_{xc0}\psi_{xc1}$, where $0, 1$ indicates QD0 and QD1.

Let us  consider the biexcitonic shift $\Delta \varepsilon$, i.e.
 the energy shift in the transition  related to the creation of a second exciton in the presence of a first one. This quantity is essential for performing,
 for example, conditional operations in quantum computational
 devices;\cite{PRLBiolatti,PRBBiolatti,DeRinaldis1,Pazy} it is then 
crucial to have a quick way to estimate  $\Delta \varepsilon$ 
in order to define the correct 
range for the structure parameters.
For the case we are analyzing, where no significant excitonic  tunneling is present between different QDs and only the Hartree term plays a relevant role,
 a good approximation for $\Delta \varepsilon$
will be the Coulomb interaction average $\Delta \varepsilon=\sum_{\alpha,\beta=e,h}\Delta\varepsilon_{\alpha 0,\beta 1}$ with
\begin{equation}\label{bxs}
\Delta \varepsilon _{\alpha 0,\beta 1}=s_{\alpha,\beta}{\langle\psi_{xc,0}\psi_{xc,1}|{e^2\over \epsilon |\vec{r_{\alpha 0}}-\vec{r_{\beta 1}}|}|\psi_{xc,0}\psi_{xc,1}\rangle}
\end{equation} with
 $s_{\alpha,\beta}=-1$ if $\alpha\ne\beta$, 1 otherwise.
$\Delta\varepsilon_{\alpha0,\beta1}$ represents the Coulomb interaction 
between particle
$\alpha$ in QD0 and particle $\beta$ in QD1 in the presence of the
other particles composing the biexciton (i.e. partially including
 correlation effects).
$\Delta\varepsilon_{\alpha0,\beta1}$ is an integral over the coordinates
of all
the four particles considered: in general in fact, due to Coulomb interaction,
 it is not possible to {\it exactly} factorize
$\psi_{xc,i}$ into  single particle components.
Let us now consider a ground state biexciton and approximate 
$\psi_{xc,i}$ with   Eq.~(\ref{excwf}).
 We stress that, being the factorization done over 
the {\it collective}  coordinates {\it internal} to the single QD, such expression {\it includes} to a 
certain extent the Coulomb interaction between  electron and 
 hole in the same QD. 
It is now  possible to integrate
analytically over most of the variables.
If we additionally approximate
the electron and hole single particle wave-functions along the $z$  direction
as
$\psi_{\alpha,i}(z_{\alpha,i})\approx
\exp\left[-(z_{\alpha,i}-
\langle z_{\alpha,i}\rangle)^2/2\lambda_{\alpha,i}^2\right]/(\sqrt{\pi}\lambda_{\alpha,i})^{1/2}$
where $\langle z_{\alpha,i}\rangle=\langle \psi_{zi}(z_{\alpha,i})|\hat{z}|\psi_{zi}(z_{\alpha,i})\rangle$ and 
$\lambda_{\alpha,i}^2\equiv 2\langle \psi_{zi}(z_{\alpha,i})|\hat{z}^2|\psi_{zi}(z_{\alpha,i})\rangle$ we obtain
\begin{eqnarray}\label{bxsfin2}
 & &|\Delta\varepsilon_{\alpha0,\beta1}|={e^2\over\epsilon}\sqrt{\tilde{D}_{\alpha0,\beta1}\over\lambda_{\alpha0}^2+\lambda_{\beta1}^2}
\int_{-\infty}^\infty dz\exp{\left[-{z^2\over\lambda_{\alpha0}^2+\lambda_{\beta1}^2}
\right]}\cdot\nonumber \\
& &\exp{[(z+\Delta z_{\alpha0,\beta1})^2\tilde{D}_{\alpha0,\beta1}]}\left\{1-\phi\left[
\sqrt{(z+\Delta z_{\alpha0,\beta1})^2\tilde{D}_{\alpha0,\beta1}}\right]\right\},
\end{eqnarray}
where $\phi(x)=(2/\sqrt{\pi})\int_0^x \exp(-t^2)dt$ is the
error function, $\Delta z_{\alpha0,\beta1}\equiv \langle z_{\alpha0}\rangle-\langle
z_{\beta1}\rangle$ and
$\tilde{D}_{\alpha0,\beta1}=D_{\alpha0}D_{\beta1}/(D_{\alpha0}+D_{\beta1})$.
Here
 $D_{e/h,i}=\left((\mu/ \hbar)\left\{B_{e/h}-
[(\omega_R-\tilde{\omega}_r)^2/ B_{h/e}]\right\}\right)_i$,
with
 $B_{e/h,i}=[\tilde{\omega}_r+\omega_R (m_{e/h}/m_{h/e})]_i$ and
$i=0,1$ indicating to which QD the involved parameters  belong. 
With the use of  Eq.~(\ref{excwf}) the twelve-dimensional integral in
 Eq.~(\ref{bxs})
has been reduced to the one-dimensional integral  in Eq.~(\ref{bxsfin2}).

Figure \ref{fig2}a shows  biexcitonic shift values associated to the system QD0+QD1 in the inset, when the
barrier width $w$ 
is varied between 2 and 4 nm and QD heights 
are  respectively 2.5 and 2.7 nm.
$\Delta\varepsilon$ is  obtained using Eq.~(\ref{bxsfin2}) (solid
line), 
and compared to the ``exact'' results in 
Ref.\onlinecite{DeRinaldis2} calculated by
direct diagonalization of the 
 fully interacting biexcitonic Hamiltonian.
As can be seen in Fig.~\ref{fig2}a, Eq.~(\ref{bxsfin2}) 
 captures most of the information: its estimates are in fact at most 7\%  from the exact values.
The curve labelled by C includes Coulomb correlation effects in the excitonic wave-functions, while the NC 
 does not. 
We stress  that, in this formulation, {\it including  Coulomb interaction
 does not imply more complex calculations}, 
since it is done by the simple substitution $\omega_r\to\tilde{\omega}_r$ (see Eq.~(\ref{om_til})). 
The precision of the results is highly affected by such corrections.

Similarly to
 the biexcitonic shift,  Coulomb interaction modifies the
transition energy in the absorption spectrum
corresponding to the creation of an exciton in a certain dot in the presence 
of  
 an electron 
(hole) in another dot.\cite{Zrenner1} In this case though, the (dipole-monopole) interaction decreases as $\sim d/w^2$ ($d$ the dipole length, roughly equal to the GaN QD height), i.e. much slower than the 
dipole-dipole interaction $\sim d^2/w^3$ characterizing the biexcitonic shift. This implies
 that,
it should be possible to 
detect such a shift (and consequently the electron (hole) presence) even if
the two dots 
are not neighbors.

Let us consider the response of an array of slightly different stacked
 GaN dots, whose height is $\sim 2.5 nm$\cite{OS} and which are  separated by barriers 
$2.5 nm$ wide.
In the hypothesis that the dot QD0 contains an electron (hole), we can calculate, by using Eq.~(\ref{bxsfin2}) and by not including Coulomb interaction in the wave functions related to QD0, the energy shift connected to the creation of an exciton in  QDN, where N=1,2,3.... is the coordination number with respect to QD0.
The system is sketch in  the inset of Fig.~\ref{fig2}b. The calculated
shift $\Delta\varepsilon_{tri}$ is plotted in  Fig.~\ref{fig2}b as a function of the distance between the centers of
QD0 and QDN. 
The coordination number of the latter dot is indicated as well.
 The curve labelled by e (h) corresponds to the presence of an electron 
(hole) in QD0.
For $N=1$, $|\Delta\varepsilon_{tri}|\sim 10 meV$, but even considering $N$ as high as 5, the energy shift is still of the order of $\sim 0.5meV$, i.e. could be resolved by laser pulses as short as 2-3 ps. The asymmetry between the ``e'' and the ``h'' curves reflects the corresponding asymmetry between electron and hole wavefunctions; the sign of $\Delta\varepsilon_{tri}$ is related to the sign of the particle in QD0.

Starting from these simple observations, we can think of a non-invasive optical
read-out device for a  memory which has been written as the 
presence (logic state 1) or the absence (logic state 0) of a charge in each QD:\cite{Zrenner}  by using a laser probe centered at the chosen QDN excitonic transition energy,
 the logic state
1 (0) will correspond to the absence (presence) of the corresponding excitonic peak in the absorption spectrum.
This scheme could be also used to measure the qubit state
 in quantum computing devices\cite{storage-nature} and could in general be
a valid
alternative to the reading done, for example, by point contacts, since it avoids charge fluctuations due to the presence of currents in the system.\cite{qpc}
In addition for far enough distances 
between the ``written''  QD0 and the ``reading'' QDN, the interaction 
 becomes proportional to the total charge inside QD0, so in principle a measure of $\Delta\varepsilon_{tri}$ could be used to count the electrons (holes) that
 have been injected in  QD0. The sign of $\Delta\varepsilon_{tri}$ would be related to the sign of the net charge present in QD0. 
The desired QDN excitonic transition can be generated by
energy selective schemes\cite{PRBBiolatti} or near field techniques. 
 A plus of the proposed device is that,
 by using a long distance 
interaction, the presence of the exciton in the reading dot would not 
perturb significantly the system in the written one.
A similar scheme can of course be  implemented in different semiconductor QDs (as for example GaAs QDs), with the caveat of using an external in-plane 
electric field to reinforce dipole-monopole interactions.\cite{GaAs} 
The main advantage of using GaN quantum dots is the strong {\it built-in} electric field 
which, on the one hand simplifies the setup, and  on the other hand will never ionize 
the trapped particles.   
Moreover  in GaN QDs Coulomb interaction
 is maximized since (i) 
this system naturally {\it aligns} the charges along the growth direction, and
(ii) due to the strength of the built-in field the wave-function spreading in the growth direction is reduced.  
It is  important to stress that, for the device to work, it is sufficient 
to have a static electric field in the {\it reading} QD, so for example 
a GaN QD layer could be grown on a different substrate. 
We finally underline that the parameters  used in our calculations are in the
reach of present experimental techniques.

In summary we have proposed
an all-optical read-out device based on long-range exciton-single charge
 interactions. The latter were calculated using our semi-analytical model 
which  allows for 
a quick and very precise estimate of few particle interactions in stacked 
 GaN quantum dots.
The device exploits the built-in electric field characteristic of GaN-based
 heterostructures but may be adapted to nanostructures based on different 
semiconductor compounds.

$~$
\newpage

\begin{figure}
  \caption{
(a) Biexcitonic shift $\Delta\varepsilon$ vs distance between nearest 
neighbors dots QD0 and QD1.
The barrier width $w$ is varied between $2\div 4 nm$. 
 $\Delta\varepsilon$ is obtained using Eq.~(\ref{bxsfin2}) (solid
line). Diamonds correspond to  the ``exact'' results in 
Ref.[9]. 
Curve ``C'' (``NC'') includes (does not include)
Coulomb corrections. Inset: Schematic view of the macromolecule QD0 + QD1. The position along the growth direction occupied by each dot is marked by couples of parallel lines.
 The electronic (light grey, $e_i$) and hole (dark grey, $h_i$) clouds are sketched as well.
\newline
(b)$\Delta\varepsilon_{tri}$ vs distance between QD0 and QDN. The barrier width is fixed, $w=2.5nm$.
The coordination number N is indicated for each point. Curve ``e'' (``h'')
corresponds to an electron (hole) in QD0. Inset:  As for inset of (a) 
but for the 
stacked-QD array QD0, ...QDN,
 when QD0 is occupied by 
an electron (left) or by a hole (right)
}
\label{fig2}
\end{figure}
$~$
\newpage

\begin{figure}
\unitlength 1cm
\begin{picture}(5.0,6.0)
\put(-10,-18){\makebox(5.0,6.0){
\includegraphics{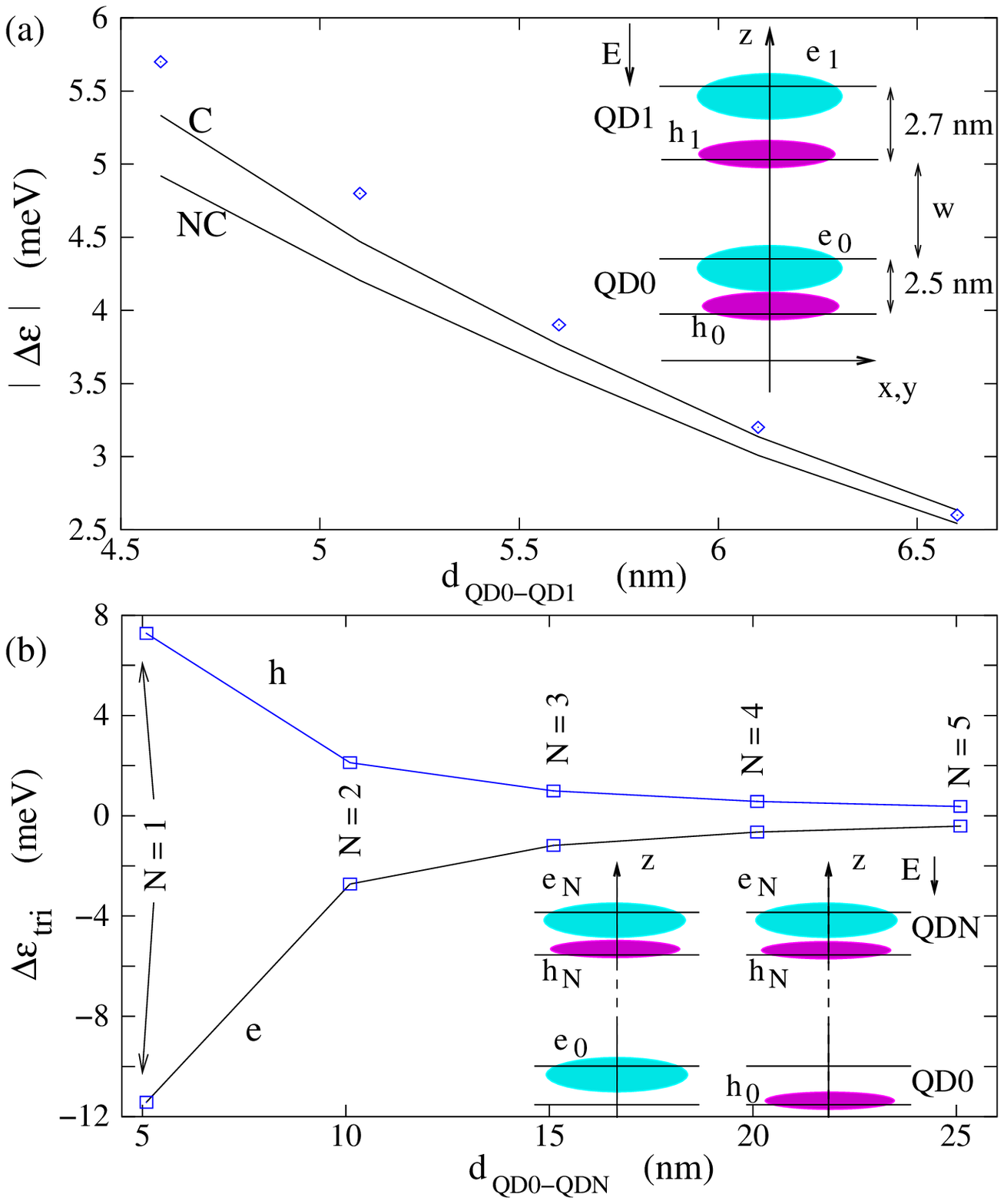}
}}
\end{picture}
 \end{figure}


\begin{thebibliography}{100}
\bibitem{wfz}The $z$-component of the wavefunction is much less affected by Coulomb correlations since (i) the particle confinement along such direction is stronger and (ii) due to the intrinsic electric field, electron and hole  wavefunctions do not significantly overlap.
\bibitem{qpc}As for point contacts, though, since both reading schemes are based on Coulomb interaction, the measure outcome could be influenced by the presence of  spurious charges in the system.
\end{thebibliography}
\end{document}